\documentclass[aps,pra,reprint,showpacs,floatfix,twocolumn]{revtex4-1}

\usepackage{amsmath,amsthm,amsfonts,amssymb,amscd,color}
\usepackage[english]{babel}
\usepackage[utf8]{inputenc}
\usepackage{graphics}
\usepackage{graphicx}
\usepackage[T1]{fontenc}
\usepackage{latexsym}
\usepackage{enumerate}
\usepackage{natbib}
\usepackage{epstopdf}
\usepackage[plainpages=false,pagebackref=false,pdftex]{hyperref}

\newcommand{\lp}{\left(}
\newcommand{\rp}{\right)}

\newcounter{eqs}
\newenvironment{eqs}{\refstepcounter{eqs}\equation}{\tag{S\theeqs}\endequation} % Muda a numeracao das equacoes para (S#)
\newcounter{eqns}
\newcommand{\ket}[1]{\ensuremath{|{#1}\rangle}} % comando para kets
\newcommand{\bra}[1]{\ensuremath{\langle{#1}|}} % comando para bra
\newcommand{\avg}[1]{\ensuremath{\langle{#1}\rangle}} % comando para médias quânticas
\newcommand{\braket}[2]{\ensuremath{\langle{#1}|{#2}\rangle}}  %comando para produto escalar entre estados
\newcommand{\braketop}[3]{\ensuremath{\langle{#1}|{#2}|{#3}\rangle}}  %comando para produto escalar com operador
\def\id{\rm 1\!\!\hskip 1 pt l}  %Unidade matricial

\begin{document}

\title{Achieving metrological precision limits through post-selection} 
\author{G. Bi\'e Alves, A. Pimentel, M. Hor-Meyll, S. P. Walborn, L. Davidovich, and R. L. de Matos Filho}
\affiliation{Instituto de F\'isica, Universidade Federal do Rio de Janeiro, P.O.Box 68528, Rio de Janeiro, RJ 21941-972, Brazil}
\begin{abstract}
Post-selection strategies have been proposed with the aim of amplifying weak signals,  which may help to overcome detection thresholds associated with technical noise in high-precision measurements. Here we use an optical setup to experimentally explore two different post-selection protocols for the estimation of a small parameter: a weak-value amplification procedure and an alternative method, that does not provide amplification, but nonetheless is shown to be more robust for the sake of parameter estimation.  Each technique leads approximately to the saturation of quantum limits for the estimation precision, expressed by the Cram\'er-Rao bound.  For both situations, we show that information on the parameter is obtained jointly from the measuring device and the post-selection statistics. \end{abstract}
\maketitle

\section{Introduction}
Reaching the ultimate precision limits in the estimation of parameters is an important challenge in science. Usually, this estimation is made by measuring the state of a probe that has undergone a parameter-dependent process. 
%Quantum mechanics imposes bounds on these estimations, due to restrictions on possible measurements and the imperfect distinguishability of states corresponding to nearby values of the parameter to be estimated. On the other hand, it is known that quantum features of the probe help to increase the precision, for the same amount of resources (which may be, for instance, the number of probes used in the estimation procedure). 
Post-selection techniques, stemming from the pioneering work of Y. Aharonov and collaborators \cite{aharonov1964time,aharonov1988result},  have been proposed with the aim of amplifying the signal obtained from the probe. In this formulation, the quantum system being analyzed gets coupled to a measuring apparatus (usually called ``meter'') through a unitary operation, which involves operators $\hat A$ for the system ${\cal A}$ and $\hat M$ for the meter ${\cal M}$, and depends on the parameter $g$ to be estimated. The goal is to estimate $g$  by measuring the change of an observable of the meter after the joint unitary evolution, given that a specified state of  ${\cal A}$ was successfully post-selected.  For a small coupling constant $g$, the shift of the mean value of the relevant meter observable is modified by a prefactor, known as the weak-value $A_w=\braketop{\psi_f}{\hat A}{\psi_i}/\braket{\psi_f}{\psi_i}$, where $\ket{\psi_i}$ and $\ket{\psi_f}$ are the initial and the post-selected states of ${\cal A}$, respectively. This quantity allows one to observe amplification effects provided the initial and the final state of the system are almost orthogonal, so long as the weak-value regime remains valid. The regime of validity of this result has been analyzed in several publications \cite{sudarshan,kofman2012nonperturbative,alves2014weak}. 

The possibility of amplifying a tiny displacement of the meter -- weak-value amplification (WVA) -- has been envisaged as a valuable resource for the estimation of the coupling constant $g$, eventually circumventing technical thresholds that may hinder the evaluation of this parameter  \cite{PhysRevA.85.060102,PhysRevLett.107.133603,PhysRevX.4.011031,PhysRevLett.105.010405}.  WVA experiments have been performed with this metrological purpose \cite{starling2009optimizing,Viza:13,viza2014experimentally}, while claiming practical advantages. Moreover, alternative protocols have been proposed \cite{PhysRevA.88.023821,PhysRevLett.113.030401} to enhance the precision of the technique. However, there has been a long debate in the literature whether this post-selection process can actually be beneficial for parameter estimation \cite{PhysRevA.85.060102,PhysRevLett.107.133603,PhysRevX.4.011031,PhysRevLett.105.010405,PhysRevA.87.012115,tanaka2013information,PhysRevLett.112.040406,combes2014quantum,knee2014weak}. Indeed, the amplification of the signal comes at the cost of discarding most of the statistical data, due to the post-selection procedure. 
   \begin{figure}[b]
\centering
\includegraphics[width=0.48\textwidth]{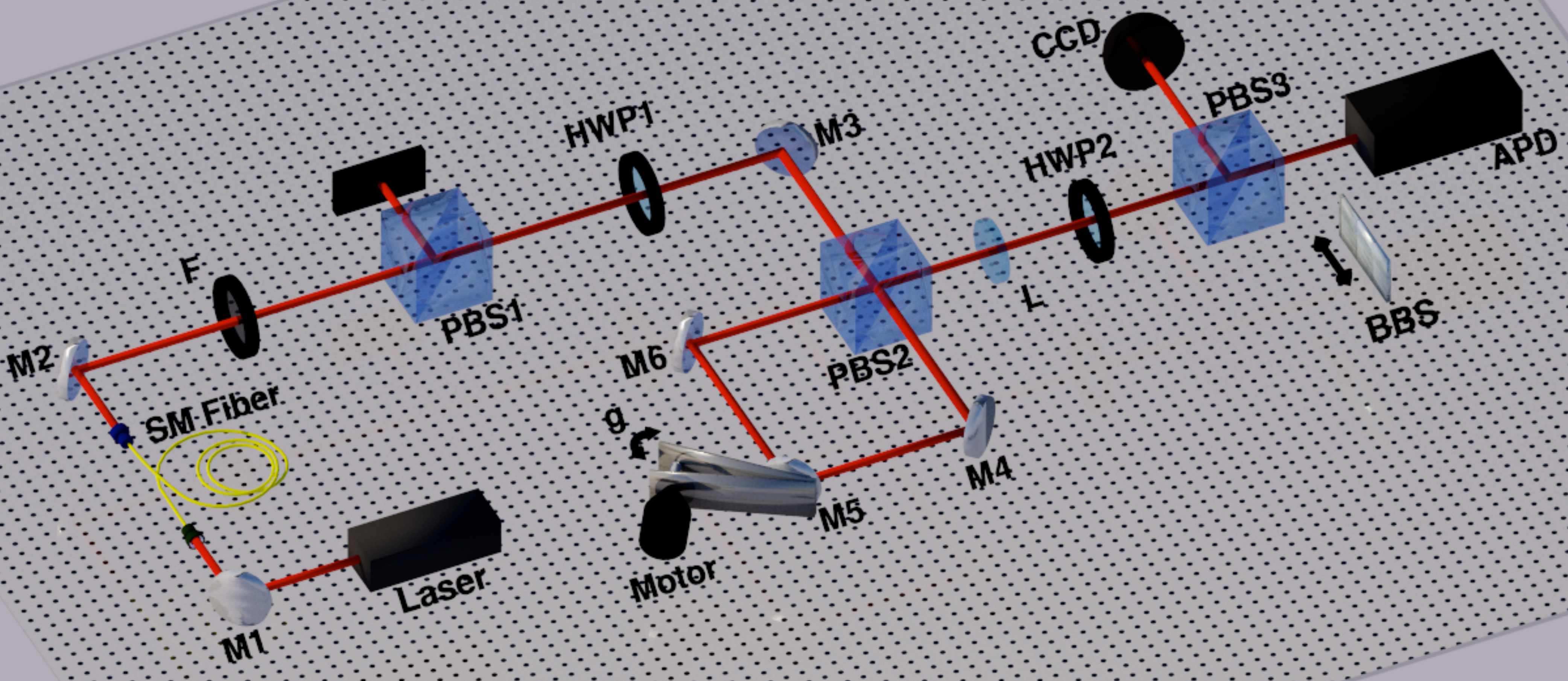}
\caption{Experimental setup. The goal is to estimate the deflection of a mirror, represented by $g$.  See text for complete description of the experiment. \label{fig:setup}}
\end{figure}

Here we experimentally investigate the estimation of  a small deflection $g$ of a mirror inside a Sagnac interferometer within the framework of quantum metrology.  We employ two post-selection protocols, which were shown \cite{alves2014weak} to lead to the ultimate quantum limits for precision, for sufficiently small $g$.  In the first one, related to the WVA approach,  we  explore the region of validity of WVA and show that, beyond this region, when the  meter does not give useful information on $g$,  estimation of this parameter can be obtained from the statistics of post-selection \cite{alves2014weak}.  We also experimentally demonstrate a post-selection procedure which, even though not leading to WVA, may also reach the fundamental limits of precision, but with a much larger post-selection probability. This implies that the number of events registered by measuring the meter is much larger than that in the WVA scheme, for the same amount of resources. This reflects in our experimental results, which clearly show that this second procedure leads to a more efficient determination of probabilities regarding the meter, in terms of frequencies of clicks in the measurement apparatus.

\section{Experimental setup}
The experimental setup is shown in Fig.~\ref{fig:setup}. A red diode laser ($\lambda = 650$ nm) is sent through a single-mode optical fiber (SM-Fiber) and decoupled by an objective lens, producing, in good approximation, a collimated free-space Gaussian beam with a width $\Delta= 286\, \mu$m. A 650$\pm$13 nm bandpass filter removes unwanted light. The polarization degree of freedom corresponds to the {\em system}, while the transverse spatial degree of freedom of the beam stands for the {\em meter}. A polarizing beam splitter (PBS1) and a half wave plate (HWP1) are used to prepare a linear-polarization state $|\psi_i\rangle$. Accordingly, the system-meter input state for the interferometer is well described by:
\begin{equation}
%|\Psi\rangle=
\!\!|\Psi_i\rangle\!\equiv\!|\psi_i\rangle\otimes |\phi_i\rangle\!=\![\cos(\theta_i/2) |H\rangle\! +\! \sin(\theta_i/2)|V\rangle]\! \otimes\! |\phi_i\rangle,
\label{eqtheta}\end{equation}
where $|H\rangle$ and $|V\rangle$ represent the horizontal  and vertical polarization states, respectively, and $|\phi_i\rangle$ stands for the initial transverse spatial state. The Sagnac interferometer is composed of three mirrors (M4, M5 and M6) and a polarizing beam splitter (PBS2). The horizontal polarization component of the input beam propagates through the interferometer in the clockwise direction, while the vertical one circulates in the counterclockwise direction, recombining again at PBS2.  A stepper motor controls the deflection angle of mirror M5. This results in transverse momentum shifts in opposite directions for the  horizontal- and vertical-polarized components, respectively. Therefore, the overall effect of the interferometer on the input beam can be represented by the unitary operator:
\begin{equation}\label{unitary}
\hat{U} = e^{-ig\hat{\sigma_3}\hat{x}}
\end{equation}
where $\hat{\sigma_3}=|H\rangle\langle H|-|V\rangle\langle V|$, $\hat{x}$ represents the transverse position operator and $g$ is the shift in transverse momentum, which is much smaller than the wavenumber $k_0$ of the light beam. After the interferometer, a $f=250\;$mm lens (L) implements a Fourier transform of the transverse spatial field at mirror M5 onto the detection plane, defined by the detection aperture of a single-photon avalanche detector (APD). The polarization measurement setup consists of a half-wave plate (HWP2) and a polarizing beam splitter (PBS3), which allows for post-selecting any linear polarization state $|\psi_f\rangle$. 
A sliding beam-blocking stage (BBS) is used for the meter measurement after post-selection. This system works like a quadrant detector.  The detection aperture of the APD is 8 mm diameter, much larger than the beam.  By counting photons while blocking half of the detector, we can determine the center of the beam, as will be discussed below.   

The Cram\'er-Rao inequality provides the lower bound on the uncertainty $\delta g$ in the estimation of the parameter $g$: $\delta g\geq 1/\sqrt{\nu F(g)}$.  Here $\nu$ is the number of repetitions of the measurement and $F(g)$ is the Fisher information, defined by $F(g)=\sum_j[1/P_j(g)][dP_j(g)/dg]^2$, where $P_j(g)$ is the probability of obtaining experimental result $j$, given that the value of the parameter is $g$.  The maximization of  $F(g)$ over all possible measurements on the system yields the \textit{quantum} Fisher information \cite{PhysRevLett.72.3439,helstrom_quantum_1976}, which provides the ultimate precision bounds. For pure initial states and unitary evolutions, it is given by $\mathcal{F}=4(\Delta\hat H)^2=4(\avg{\hat H^2}-\avg{\hat H}^2)$, where $\hat H$ is the generator of the unitary transformation, and the averages are taken with respect to the initial quantum state. From Eq.~\eqref{unitary}, $\hat H=\sigma_3\hat x$, so, assuming that initially $\avg{\hat x}=0$ (balanced-meter condition), one has  $\mathcal{F}=4\avg{\hat x^2}$, where the average is taken in the initial state $|\phi_i\rangle$ of the meter. Therefore, the larger the variance of position in the initial state of the meter,  the more information about the parameter is imprinted by the unitary evolution on the state of system+meter.  

Under post-selection on a state $|\psi_f\rangle$ of the system, the Fisher information about $g$ can be decomposed as \cite{alves2014weak,PhysRevLett.114.210801} $F_{ps}(g)=p_f(g)F_m(g)+F_{p_f}(g)$, where $F_m(g)$  is the Fisher information associated to measurements on the state of the meter after post-selection, and $p_f(g)$ is the probability that the post-selection succeeds, while
\begin{equation}\label{fam}
F_{p_f}(g)=\frac{1}{p_f(g)[1-p_f(g)]}\left[\frac{dp_f(g)}{dg}\right]^2 
\end{equation}
is the Fisher information on $g$ corresponding to $p_f(g)$. 

In \cite{alves2014weak}, it was shown that, for sufficiently small $g$, and for optimal measurement on the meter, post-selection on either the state $\ket{\psi_f}=\hat A\ket{\psi_i}/\sqrt{\avg{\hat A^2}}$ or $|\psi_f\rangle=|\psi_i\rangle$ leads to a value of $F_{ps}$ that coincides with the quantum Fisher information ${\cal F}(g)$, up to terms of $O(g^2)$. Therefore, under these conditions, this procedure yields optimal information on the parameter. These results, developed in \cite{alves2014weak}, differ from the standard WVA approach in two important features: (i) the best post-selection \emph{is not} on a state of ${\cal A}$ quasi-orthogonal to the initial state; and (ii) one should consider, in general, the statistics of post-selection,  in addition to the results stemming from measurements on the meter. The choice $|\psi_f\rangle=|\psi_i\rangle$ does not lead to WVA, but implies a probability of post-selection much higher than the WVA procedure. We will show in this paper that this results in a more reliable determination of the probability distribution of the displacement of the meter, which is used for the estimation of $g$. 

\section{Experimental procedure}

The experiment consists in applying a small misalignment $g$ to mirror M5, sending light in the state $|\psi_i\rangle\otimes|\phi_i\rangle$ into the interferometer, recording the statistics of post-selection events on the polarisation state $|\psi_f\rangle$, and measuring the mean transverse momentum of the light beam, after successful post-selection on $|\psi_f\rangle$. From the data, one finally extracts an estimation for $g$, using a maximum likelihood estimator.
The post-selection probability $p_f(g)$ is obtained by measuring, without the BBS,  the photon counts $N_f$ reaching the APD when the setup is adjusted for post-selecting the state $|\psi_f\rangle$, as well as the photon counts $N_f^\bot$ corresponding to the state orthogonal to $|\psi_f\rangle$. For a large number of counts, $p_f(g)$ is given by:
\begin{equation}
p_f(g)=\frac{N_f(g)}{N_f(g)+N_f^\bot(g)}\, .
\end{equation}
The corresponding theoretical model leads to (see Appendix \ref{appx-pf})
\begin{equation}\label{pf-D}
p_f(g)=\frac{1}{2}(1+\nu_0\cos^2\theta_i \pm\nu_{\pi/2}\sin^2\theta_i\,e^{-2g^2\Delta^2})\,,
\end{equation}
where the $+(-)$ sign corresponds to the post-selection $\ket{\psi_f}=\ket{\psi_i}$ ($\ket{\psi_f}=\hat\sigma_3\ket{\psi_i}$) and $\nu_{\pi/2}$ is the visibility of the interferometer and $\nu_0$ is related to the extinction ratio of the polarization optics.In our experiment, $\nu_0=0.998$ and $\nu_{\pi/2}=0.966$.

All of the information about $g$ that is encoded in the meter state $|\phi_f\rangle$ can be retrieved via  an optimal projective measurement. As described in \cite{alves2014weak}, measurement of the observable $\hat W=\hat k$, conjugate to $\hat M=\hat x$, is optimal for both post-selection procedures, as long as the meter state remains Gaussian after the post-selection. In fact, in the region of validity of the WVA, the distribution of the eigenvalues of the observable $\hat k$ in the final meter state remains Gaussian, with the same variance as in the initial state and shifted mean~\cite{aharonov1988result,alves2014weak}. In this situation, measurement of the mean value $\avg{\hat k}$ is equivalent to measurement of $\hat k$. 
At the focal plane of lens L, a shift in the mean transverse momentum $\avg{\hat k}$ of the beam at mirror M5 is converted into a displacement of the beam center, given by
\begin{equation}\label{displacement-lens}
d=f\avg{\hat k}/k_0\, ,
\end{equation}
where $k_0 = 2\pi/\lambda$, while the beam is resized to $\Delta_f=f/2k_0\Delta$.  For this reason, in order to measure a shift in $\avg{\hat k}$, the beam center is measured with the BBS positioned at the focal plane of  lens L. Information about the beam displacement is then obtained via measurement of the number of photons reaching the two halves of the transverse plane during a given sampling time, which is equivalent to a split detector technique \cite{putman1992,barnett_ultimate_2003}.  For a Gaussian beam with diameter $\Delta_f$ (at the focus), subjected to a displacement $d$, we have
\begin{equation}\label{split-detection}
\frac{|N_R-N_L|}{N_R+N_L}=\sqrt{\frac{2}{\pi}}\frac{d}{\Delta_f}\, ,
\end{equation}
where $N_L$($N_R$) is the number of photons detected on the left (right) half-plane. 
%The sampling time was $\Delta t=0.1\,s$, giving an average of $N\approx 100000$ photons.
The corresponding theoretical model, taking into account the imperfect visibility, yields for the mean momentum deflection (Appendix \ref{appx-meter})
\begin{equation}
\avg{\hat k}=\frac{-2g\cos\theta_i}{(1+\nu_0\cos^2\theta_i\pm\nu_{\pi/2}\sin^2\theta_i\,e^{-2g^2\Delta^2})} \label{shift-k-D}\,.
\end{equation}

\section{Experimental results}
 
We send on average $N\approx 10^5$ photons into the interferometer and determine $p_f(g)$ and $\langle \hat{k} \rangle$ as described above. From the measured data we obtain an estimate for $g$ via a maximum likelihood estimator (Appendix \ref{appx-MLE}). In order to access the precision of the estimation procedure and compare it with the theoretical bounds, we repeated this measurement process $100$ times and used the variance of the resulting estimates of $g$ as the uncertainty in our estimation procedure.

In Appendix \ref{appx-exp}, it is shown that the experimental data are in good agreement with the proposed theoretical model for the post-selection probability, as given by Eq.~\eqref{pf-D}, as well as for the mean beam displacement, as given by  Eq.~\eqref{displacement-lens} and Eq.~\eqref{shift-k-D}, for both types of post-selection strategies.
\begin{figure}[t]
\centering
\includegraphics[width=0.47\textwidth]{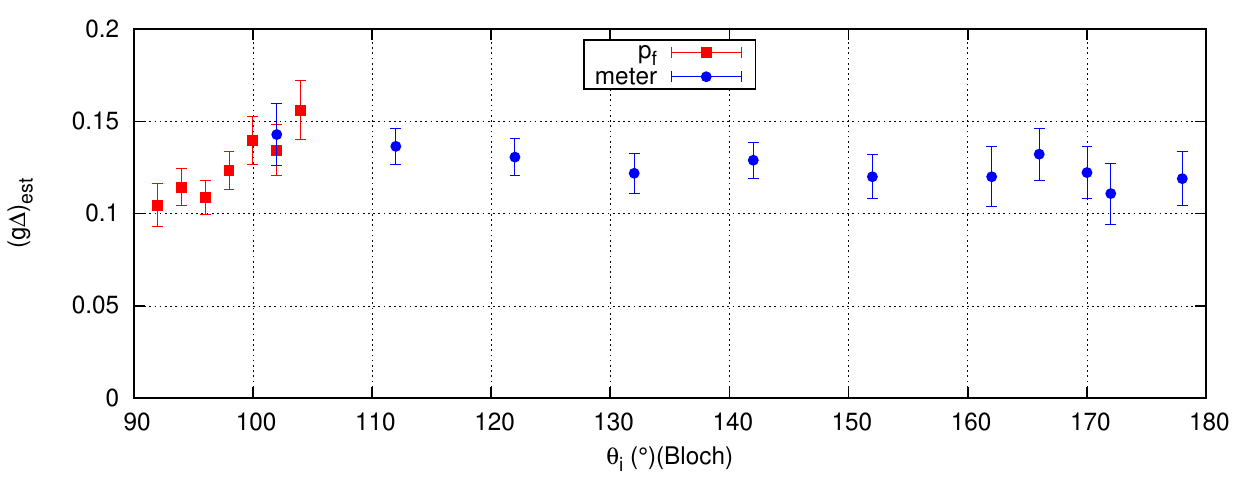}
\caption{Estimated value of $g\Delta$ from the experiment, for post-selection in $|\psi_f\rangle=\hat\sigma_3 |\psi_i\rangle$. Blue circles correspond to estimates from measurements on the meter while red squares correspond to estimates obtained via the statistics of post-selection events.  Error bars represent 3-$\sigma$ dispersion. \label{fig:g-apsi}}
\end{figure}

Figure~\ref{fig:g-apsi} shows the estimated values of the dimensionless parameter $g\Delta$ as function of the angle $\theta_i$ that defines the initial state $|\psi_i\rangle$,  for the post-selection procedure related to the WVA, where $|\psi_f\rangle=\hat\sigma_3 |\psi_i\rangle$. Notice that, in contrast to traditional WVA approaches, the post-selection is not, in general, in a state quasi-orthogonal to $|\psi_i\rangle$.    As $|\psi_f\rangle$ approached orthogonality to $|\psi_i\rangle$, we could not provide reliable estimates for $g\Delta$ based on measurements on the meter alone.  In our experiment this corresponds to the region $90^\circ \lesssim \theta_i  \lesssim 100^\circ$, due to the value of $g \Delta$ used in the experiment.  There are both fundamental and practical reasons for this.  First, the wave packet of $|\phi_f\rangle$ begins to distort and lose its Gaussian shape.  As a result, direct measurement of the mean value $\avg{\hat k}$ is no longer optimal, and the the split detector is unable to measure the displacement of the non-gaussian beam correctly.  Finally,  the quantity of information about $g$ encoded in the meter drops sharply to zero. However, in exactly this region where there is almost no information about the parameter $g$ in the state of the meter and the WVA approach is no longer applicable, we have obtained excellent estimates of $g\Delta$ by considering only the post-selection probability $p_f(g)$.  Thus, by taking into account information from both the meter and the post-selection statistics, we are able to provide consistent estimates of $g\Delta$ for all values of $\theta_i$.

\begin{figure}
\centering
\includegraphics[width=0.47\textwidth]{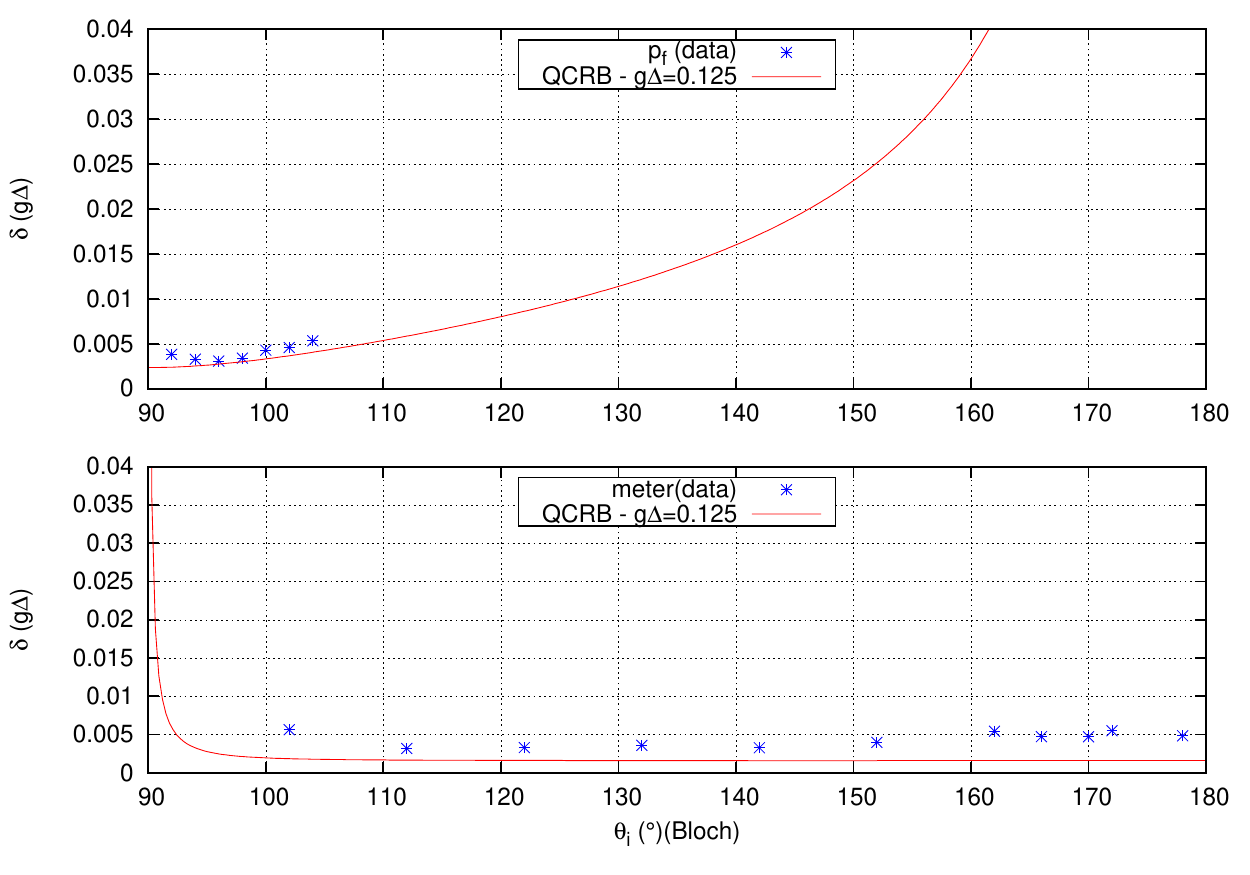}
\caption{Comparison between theoretical and experimental uncertainties for estimation of the the dimensionless parameter $g\Delta$ as function of the initial state $|\psi_i\rangle$, for the post-selection $|\psi_f\rangle=\sigma_3|\psi_i\rangle$. The measured uncertainties (stars) are shown for estimation via the statistics of post-selection events (top figure) and measurement on the meter state (bottom figure). The lines show the corresponding  quantum bounds, taking into account the imperfections in the interferometer.\label{fig:compare-qcrb-apsi}}
\end{figure}

In Fig.~\ref{fig:compare-qcrb-apsi} we compare the experimental uncertainties in the estimation of the dimensionless parameter $g\Delta$ with the corresponding quantum Cram\'er Rao bounds (QCRB), where we have taken into account the imperfect visibility of the interferometer. No other experimental imperfections are considered.   This figure shows that our approach, based on the prevalence of the meter and the post-selection statistics in different regions, leads to uncertainties very close to the theoretical bounds.

It is important to notice also that the mean number of photons used to estimate $\avg{\hat k}$ from the meter decreases steadily as $\theta_i$ decreases from $180^\circ$  to $90^\circ$ ($\langle \psi_f|\psi_i\rangle \rightarrow 0$). Since the maximum likelihood estimator is only asymptotically consistent, the reduction in the number of photons used in the estimation of the parameter $g\Delta$ via measurement on $|\phi_f\rangle$  impairs the performance of the estimator, increasing its biasness and uncertainty.  This is a drawback of post-selection procedures related to WVA when compared to other strategies, if the total resources used in the experiment (photon number in our case)  are kept constant.

\begin{figure}[t]
\centering
\includegraphics[width=0.47\textwidth]{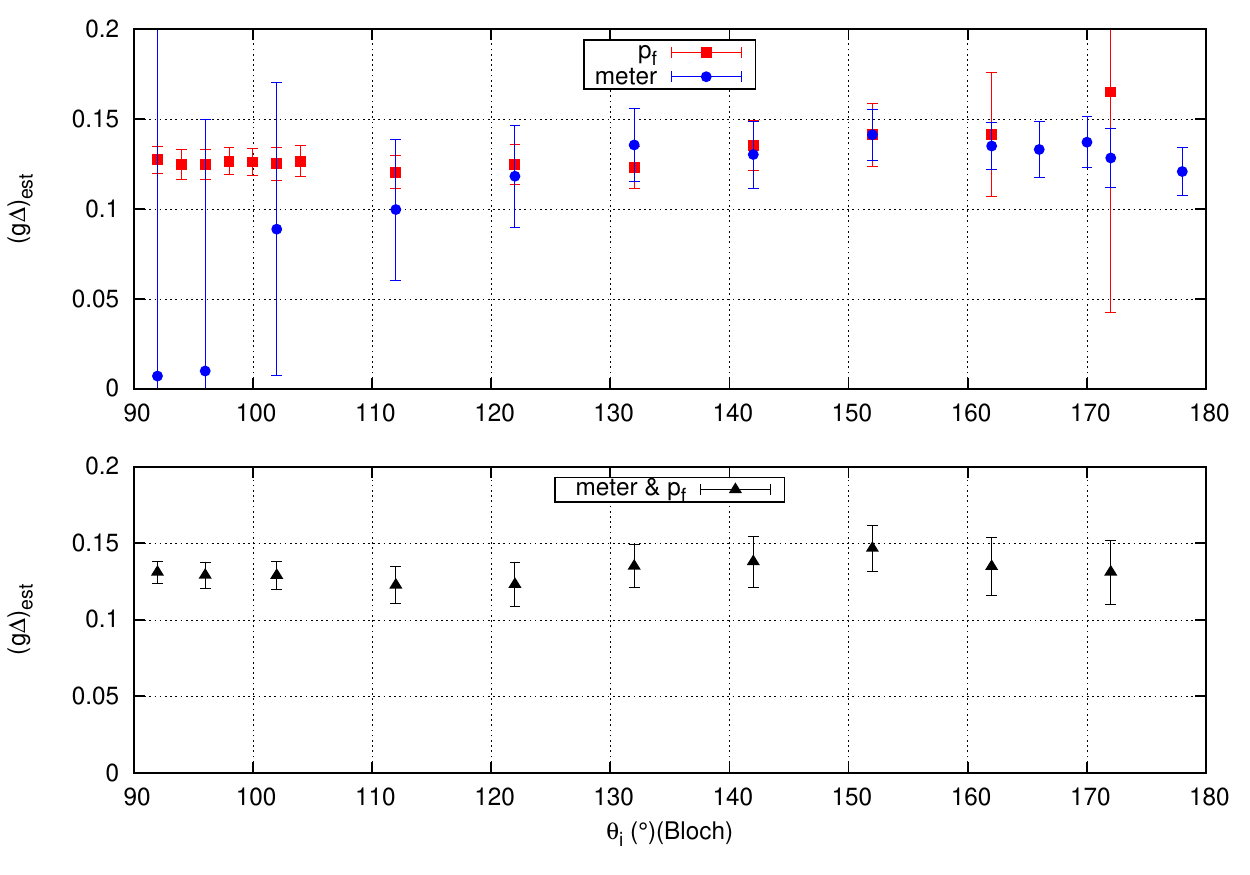}
\caption{ Estimated value of $g\Delta$ from the experiment, for post-selection in $|\psi_f\rangle=|\psi_i\rangle$. Blue circles correspond to estimates from measurements on the meter, red squares correspond to estimates obtained via the statistics of post-selection events while black triangles correspond to estimates using information of both the meter and the statistics of post-selection events. Error bars represent 3-$\sigma$ dispersion. \label{fig:g-psi}}
\end{figure}
\begin{figure}[t]
\centering
\includegraphics[width=0.47\textwidth]{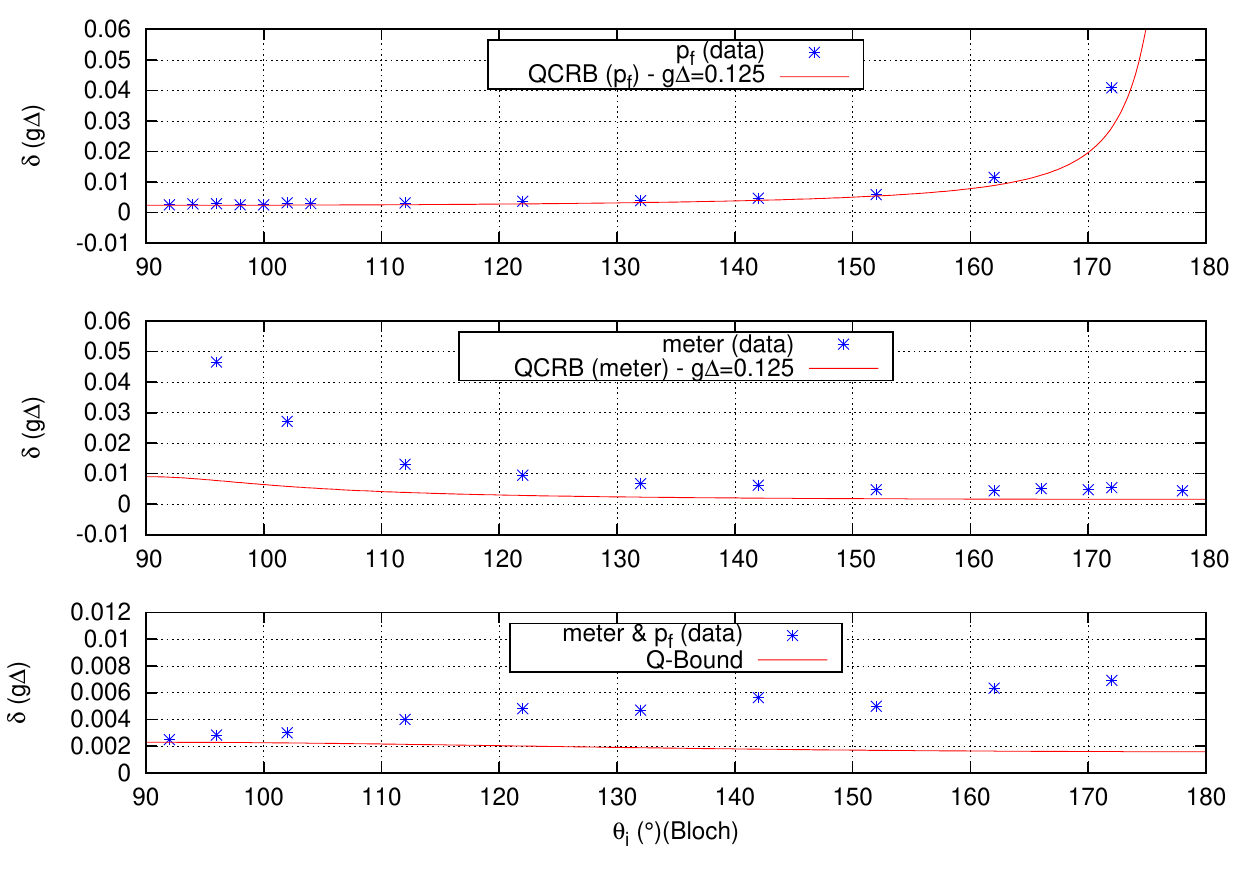}
\caption{Comparison between theoretical and experimental uncertainties for estimation of the dimensionless parameter $g\Delta$ as function of the initial-state parameter $\theta_i$, for the post-selection $|\psi_f\rangle=|\psi_i\rangle$. The measured uncertainties (blue stars) are shown for estimation via the statistics of post-selection events (top figure), measurement on the meter state (middle figure) and by using information of both the meter and the statistics of post-selection events (bottom figure). The lines show the corresponding quantum bounds on precision (see text). The scale of the bottom figure is highly amplified. \label{fig:compare-qcrb-psi}}
\end{figure}

We consider now the second post-selection strategy  which relies on post-selection onto  $|\psi_f\rangle=|\psi_i\rangle$. Fig.~\ref{fig:g-psi} shows the estimated values of $g\Delta$ as function of the initial state $|\psi_i\rangle$ for this case. Contrary to the first case, here information about the parameter $g$ is distributed among the meter and the post-selection statistics for almost the whole range of $\theta_i$ values~\cite{alves2014weak}. As a result, the best estimation of the parameter $g\Delta$ is provided by using data from both the meter and the statistics of post-selection events.
Information about $g$ encoded in the statistics of post-selection events begins to decrease for $\theta_i\approx 160^\circ$, and is zero for $\theta=180^\circ$. This explains the increase in the error bars of the corresponding estimates for this range of values of $\theta_i$.  On the other side, information about $g$ encoded in the meter dwindles when $\theta_i$ decreases from $120^\circ$ to $90^\circ$, but never goes to zero. The degradation of the corresponding estimates in the region between $110^\circ$ and $90^\circ$ is due to the fact that measuring   $\avg{\hat k}$ is no longer optimal and, in fact, the information about $g$ extractable via $\avg{\hat k}$ does goes to zero as the value of $\theta_i$ approaches $90^\circ$.  This behaviour is  clearly displayed  in Fig.~\ref{fig:compare-qcrb-psi}, which compares the experimental uncertainties in the estimation of the dimensionless parameter $g\Delta$. As before, the QCRB takes into account the slighlty reduced visibility of the interferometer, and no other experimental imperfection. Figure \ref{fig:compare-qcrb-psi} also shows that the experimental uncertainty in the estimates produced from information of both the meter and the statistics of post-selection events are very close to the theoretical quantum bounds for all values of $\theta_i$. 

It is important to notice that, contrary to the WVA post-selection procedure,  the probability of successful projection onto $|\psi_f\rangle=|\psi_i\rangle$ is close to unity for the whole range of values $\theta_i$. Thus,  the mean number of photons used to estimate $\avg{\hat{k}}$ from the meter remains very close to the total number of photons injected into the interferometer, and does not lead to the degradation of the performance of the maximum likelihood estimator, as was the case for the WVA-related strategy. Furthermore,  the Fisher information corresponding to $p_f(g)$ is now relevant over a wider range of initial states, as compared to the previous post-selection scheme, which is yet another advantage of the present procedure, since measuring $p_f(g)$ is simple to implement  and is always optimal, independently of the initial state. This is in stark contrast to optimal measurements on the meter, which depend on the initial state, and are therefore more challenging to implement.   These two facts lead to the consistency of the estimates for any initial state $|\psi_i\rangle$, as shown in Fig.~\ref{fig:g-psi}, which result in a highly effective and robust metrological protocol that can, in principle, reach the ultimate precision bounds on parameter estimation.  

\section{Conclusion}

We have experimentally investigated two post-selection-based measurement procedures from a quantum metrology point of view.  One of these is a new post-selection procedure, which presents considerable advantages over the weak-value amplification (WVA) scheme. The first method considered in this paper, related to the WVA procedure, fails to provide a reasonable estimation of the parameter (tilt of a mirror in the interferometer, described by $g$) from the state of the meter alone when the post-selection is around the region of highest amplification. In this same region, the non-Gaussian profile of the wave packet describing the meter makes it difficult to implement an optimal measurement on it. Furthermore, the reduction in the number of photons that remain, due to the post-selection on a nearly orthogonal state,   impairs the performance of the corresponding estimator. We show that accurate estimation of $g$ can be performed in this case when the statistics of the post-selection of the system are taken into account.    In the new method, there is almost no reduction in detection events due to post-selection, since the corresponding probability remains close to one for all initial states.  This leads to better performance of the meter estimator. In addition, information on $g$ is now distributed between the meter and the post-selection statistics over a wider range of initial states, which allows one to benefit from the information encoded in the post-selection statistics, which is simple to obtain experimentally from a measurement procedure that is always optimal, regardless of the initial state. This method leads to uncertainties in the estimation of $g$ that are closer to the quantum Cram\'er-Rao bound, particularly in the region where the information from the post-selection statistics is dominant. 

Our experiment serves as a proof-of-principle demonstration that post-selection protocols can be metrologically efficient, as long as the information encoded in the post-selection statistics is also taken into account.  It also throws new light on the subtle connection between post-selection procedures and quantum metrology, offering a viable and easy-to-implement procedure  that can be easily generalized to other parameter estimation  tasks. 

\section{Acknowledgments}
This research was supported by the Brazilian agencies FAPERJ, CNPq, CAPES, and the National Institute of Science and Technology for Quantum Information.

\newpage
\appendix
\onecolumngrid

\section{Adapting the theoretical model to the experimental conditions}

\subsection{The post-selection probability}\label{appx-pf}

As described in the main text, the expected interaction implemented between the system and the meter by the Sagnac interferometer would correspond to an unitary operator $\hat U_g=e^{-ig\hat\sigma_3\hat x}$, as in Eq.(2). This leads to a probability of post-selection given by
\begin{eqs}\label{pf}
p_f(g)=|\braketop{\psi_f}{\hat U}{\psi_i}\ket{\phi_i}|^2=\frac{1}{2}(1+\cos^2\theta\pm e^{-2g^2\Delta^2}\sin^2\theta)\,,
\end{eqs}
where the +(-) sign correspond to the post-selected state $\ket{\psi_f}=\ket{\psi_i}$ ($\ket{\psi_f}=\hat\sigma_3\ket{\psi_i}$). However, one can easily seen that, before the interaction takes place ($g=0$) the post-selection probability onto the same initial state ($\ket{\psi_f}=\ket{\psi_i}$) is $1$ irrespective of the initial state on the Bloch sphere. This implies that the visibility of the interferometer should be 100\%, whatever the initial state is, which does not correspond to the real experimental conditions, where tiny unwanted misalignment of the interferometer may affect the quality of the interference.

To account for a small relative misalignment in the $y$ direction (perpendicular to $x$, in the transverse plane), it is enough to consider the same interaction, but now being implemented by $\hat U_w=e^{-iw\hat\sigma_3\hat y}$. Thus, the complete evolution would be described by:
\begin{eqs}\label{evolucao-desalinhamento}
\hat U_w\hat U_g\ket{\psi_i}\ket{\phi_i}\ket{\varphi_i}\,,
\end{eqs}
where it is supposed that the spacial state is a product state between the directions $x$ and $y$, represented by $\ket{\phi_i}$ and $\ket{\varphi_i}$.

The probability of post-selection is then given by
\begin{eqs}\label{pf-desalinhamento}
\begin{split}
p_f(g)&={\rm Tr}\left( \hat P_f \hat U_w\hat U_g\ket{\psi_i}\bra{\psi_i}\otimes\ket{\phi_i}\bra{\phi_i}\otimes\ket{\varphi_i}\bra{\varphi_i}\hat U_w^\dagger\hat U_g^\dagger\right) \\
&={\rm Tr_x}\left[ \hat U_g\ket{\psi_i}\bra{\psi_i}\otimes\ket{\phi_i}\bra{\phi_i}\otimes\hat U_g^\dagger\,\,{\rm Tr_y}\lp \ket{\varphi_i}\bra{\varphi_i}\hat U_w^\dagger\hat P_f \hat U_w \rp \right]\,,
\end{split}
\end{eqs}
where $\hat{P}_f=\ket{\psi_f}\bra{\psi_f}$.
However, we have that
\begin{eqs}\label{canal-defasagem1}
\begin{split}
&{\rm Tr_y}\lp \ket{\varphi_i}\bra{\varphi_i}\hat U_w^\dagger\hat P_f \hat U_w \rp = \bra{\varphi_i}\hat U_w^\dagger\hat P_f \hat U_w\ket{\varphi_i} \\
&=\hat P_f\lp \int dy|\varphi(y)|^2\cos^2(wy) \rp +i[\hat P_f, \hat\sigma_3]\lp \int dy |\varphi(y)|^2\sin(wy)\cos(wy) \rp + \hat\sigma_3\hat P_f\hat\sigma_3 \lp\int dy |\varphi(y)|^2\sin^2(wy)\rp \\
&= (1-p/2)\hat P_f + p/2\,\hat\sigma_3\hat P_f\hat\sigma_3\,,
\end{split}
\end{eqs}
where $p/2\equiv\int dy|\varphi(y)|^2\sin^2(wy)$, since the wave-function is gaussian and the second term in Eq.\eqref{canal-defasagem1} vanishes. This implies that the evolution under the interaction $\hat U_w$ leads to a  \emph{dephasing channel} when monitoring only the $x$ direction:
\begin{eqs}
p_f(g)={\rm Tr_x}\sum_{\mu}\lp \hat P_f \hat K_\mu \hat U_g\ket{\psi_i}\bra{\psi_i}\otimes\ket{\phi_i}\bra{\phi_i}\otimes\hat U_g^\dagger \hat K_\mu^\dagger \rp \equiv {\rm Tr_x}\lp\hat P_f\hat\rho'_{S,M}(g)\rp\,,
\end{eqs}
with $\hat K_1=\sqrt{1-p/2}\,\hat\id$, $\hat K_2=\sqrt{p/2}\,\hat\sigma_3$ -- the Kraus operators -- and $\hat\rho'_{S,M}(g)=\sum_\mu K_\mu \hat U_g\ket{\psi_i}\bra{\psi_i}\otimes\ket{\phi_i}\bra{\phi_i}\otimes\hat U_g^\dagger K_\mu^\dagger$.

Besides this effect, it is assumed in the theoretical model that the polarization state $\ket{\phi_i}$ is prepared with $100\%$ efficiency. To account for partial coherence in the preparation of the system state, we add an orthogonal component in the density matrix of the original state:
\begin{eqs}\label{despolarizacao}
\hat\rho_S^i=(1-\epsilon)\ket{\psi_i}\bra{\psi_i}+\epsilon\ket{\psi_i^\perp}\bra{\psi_i^\perp}\,,
\end{eqs}
where, $\epsilon$ is expected to be very small. Together with the dephasing channel, this leads to an evolution described by
\begin{eqs}\label{evolucao-DD}
\hat\rho_{S,M}^i=\hat\rho_S^i\otimes\ket{\phi_i}\bra{\phi_i}\mapsto\hat\rho_{S,M}''(g)=\sum_\mu \hat K_\mu\hat U_g(\hat\rho_S^i\otimes\ket{\phi_i}\bra{\phi_i})\hat U_g^\dagger\hat K_\mu^\dagger\,.
\end{eqs}
After some straightforward calculation, the probability of post-selection is given by
\begin{eqs}\label{pf-DD}
\begin{split}
p_f(g)&=\sum_\mu{\rm Tr}(\hat P_f\hat K_\mu\hat U_g(\hat\rho_S^i\otimes\ket{\phi_i}\bra{\phi_i})\hat U_g^\dagger\hat K_\mu^\dagger) \\
&=\frac{1}{2}\left[1+(1-2\epsilon)\cos^2(\theta_i)\pm(1-2\epsilon)(1-p)\sin^2(\theta_i)\,e^{-2g^2\Delta^2}\right]\,,
\end{split}
\end{eqs}
where the sign $+(-)$ corresponds to a post-selection onto $\ket{\psi_f}=\ket{\psi_i}$ ($\ket{\psi_f}=\hat\sigma_3\ket{\psi_i}$). Defining the visibility for a given prepared initial state $\ket{\psi_i}$ -- before the interaction takes place -- as
\begin{eqs}
\nu_\theta=\frac{N_f(\theta)-N_f^\perp(\theta)}{N_f(\theta)+N_f^\perp(\theta)}=2p_f(0)\Big\vert_{\theta}-1\,,
\end{eqs}
we can readily interpret the parameters $\epsilon$ and $p$ in terms of interference visibility analyzed in different polarization bases (characterized by $\theta$), such that the expression for the probability can be rewritten as:
\begin{eqs}\label{pf-DD-0}
p_f(g)=\frac{1}{2}\left[1+\nu_{0}\cos^2(\theta_i)\pm\nu_{\pi/2}\sin^2(\theta_i)\,e^{-2g^2\Delta^2}\right]\,.
\end{eqs}

\subsection{The meter analysis}\label{appx-meter}

Since our purpose is to measure the mean momentum shift in the transverse plane after the post-selection (as long as the meter state remains gaussian), we require expressions for theoretical expected value of the operator $\hat k$, where $[\hat x,\hat k]=i$. Taking into account the decoherence channels presented in the last section, the meter state after the post-selection is given by
\begin{eqs}\label{meter-f-DD}
\hat\rho''_f(g)=\frac{\sum_\mu{\rm Tr}_S\lp\hat P_f\hat K_\mu\hat U_g\,\hat\rho_S^i\otimes\ket{\phi_i}\bra{\phi_i}\,\hat U_g^\dagger\hat K_\mu^\dagger\rp}{p_f(g)}\,,
\end{eqs}
where $\rm Tr_S(\cdot)$ is the trace over the system (polarization) space. The measurement of $\hat k$ is then given by
\begin{eqs}
\avg{\hat k}={\rm Tr}_M(\hat\rho''_f(g)\,\hat k)=\frac{\sum_\mu{\rm Tr}\lp \hat k \hat P_f\hat K_\mu\hat U_g\,\hat\rho_S^i\otimes\ket{\phi_i}\bra{\phi_i}\,\hat U_g^\dagger\hat K_\mu^\dagger \rp}{p_f(g)}\,.
\end{eqs}
After some straightforward calculation, we have:
\begin{eqs}\label{deslocamento-k-DD}
\avg{\hat k}=\frac{-g(\nu_0+1)\cos\theta_i}{(1+\nu_0\cos^2\theta_i\pm\nu_{\pi/2}\sin^2\theta_i\,e^{-2g^2\Delta^2})}\,,
\end{eqs}
where the sign $+(-)$ corresponds to $\ket{\psi_f}=\ket{\psi_i}$ ($\ket{\psi_f}=\hat\sigma_3\ket{\psi_i}$).

\subsection{Maximum likelihood estimation}\label{appx-MLE}

Here we describe how we provide estimatives for the desired parameter $g$ for each experimental measurement outcome. The maximum likelihood estimation procedures consists of finding the value of the coupling $g$ that best matches a given experimental result in terms of the probability of occurrence. Thus, the \emph{estimator} for $g$ is found to be the one that maximizes the theoretical probability associated with a certain measured outcome. For the case of estimation based solely on the post-selection probability $p_f(g)$, this procedure leads to solving Eq.(4) in the main text for $g$. Analogously, for the estimation based on the meter measurements, the equation to solve is given by Eq.(7) of the main text, with the aid of Eq.\eqref{deslocamento-k-DD}. However, for the estimation based on both results, the outcome is defined by the set of numbers $\{N_R,N_L,N^\perp_f\}$, where $N_R+N_L=N_f$. The likelihood probability is then given by
\begin{eqs}\label{likelihood}
\mathcal{L}=P_R(g)^{N_R}P_L(g)^{N_L}(1-p_f(g))^{N^\perp_f}
\end{eqs}
where $P_L(g)$, $P_R(g)$ are the theoretical probabilities of the meter to be detected at the left, right half of the detector. The estimator $g_{est}$ is then found by solving
\begin{eqs}\label{MLE}
\frac{\partial\mathcal{\ln L}}{\partial g}\Big\vert_{g_{est}}=0.
\end{eqs}
For the case of post-selection $\ket{\psi_f}=\ket{\psi_i}$, the meter remains approximately gaussian and the probabilities $P_{L,R}(g)$ can be calculated as
\begin{eqs}\label{P-LR}
P_R(g)=p_f(g)-P_L(g)\approx\left[\frac{1}{2}+\frac{d}{\sqrt{2\pi}\Delta_f}\right]p_f(g)\,,
\end{eqs}
where $d=f\avg{\hat k}/k_0$. Using the equations \eqref{likelihood}, \eqref{P-LR}, \eqref{deslocamento-k-DD} and \eqref{pf-DD-0} one can finally solve Eq.\eqref{MLE} by numerical methods, once there is no analytical solution.

\section{Experimental details}\label{appx-exp}

\subsection{Experimental procedure}

As explained, besides the desired parameter $g\Delta$ (to be estimated), the model incorporates the visibilities $\nu_0$ and $\nu_{\pi/2}$, which are measured before the mirror angle is displaced ($g=0$). Experimentally, the interferometer is set to the best possible alignment conditions, and the visibilities are measured when preparing the states $\ket{\psi_i}=\ket{H}$ and $\ket{\psi_i}=\ket{+}=(\ket{H}+\ket{V}/\sqrt{2})$, subtracting photocounts due to ambient noise. For our alignment conditions, we obtained $\nu_{\pi/2}=0.966$ and $\nu_0=0.998$, where this last one was expected to be very close to unity, provided the high efficiency in the wave plates.

\begin{figure}[h]
\begin{tabular}{cc}
\includegraphics[scale=0.7]{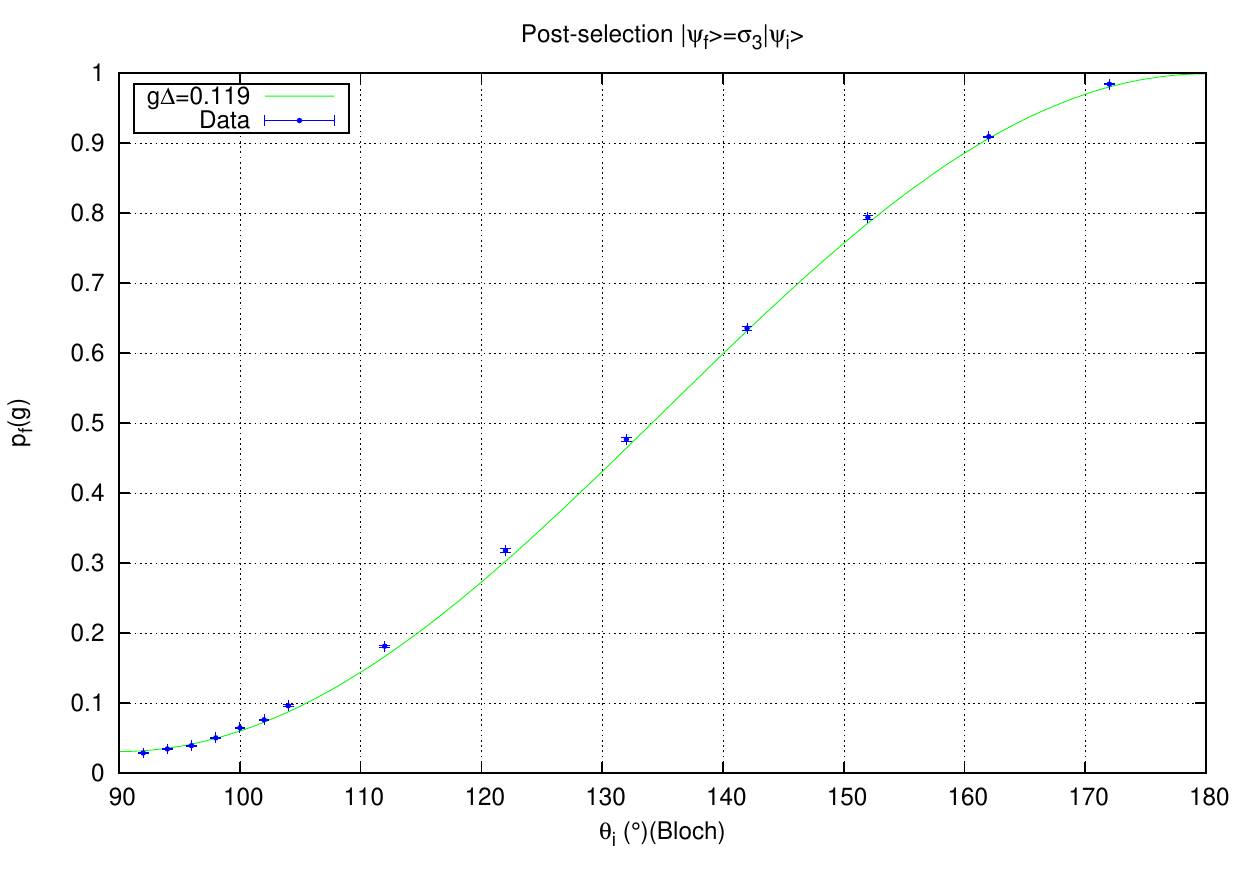}
&
\includegraphics[scale=0.7]{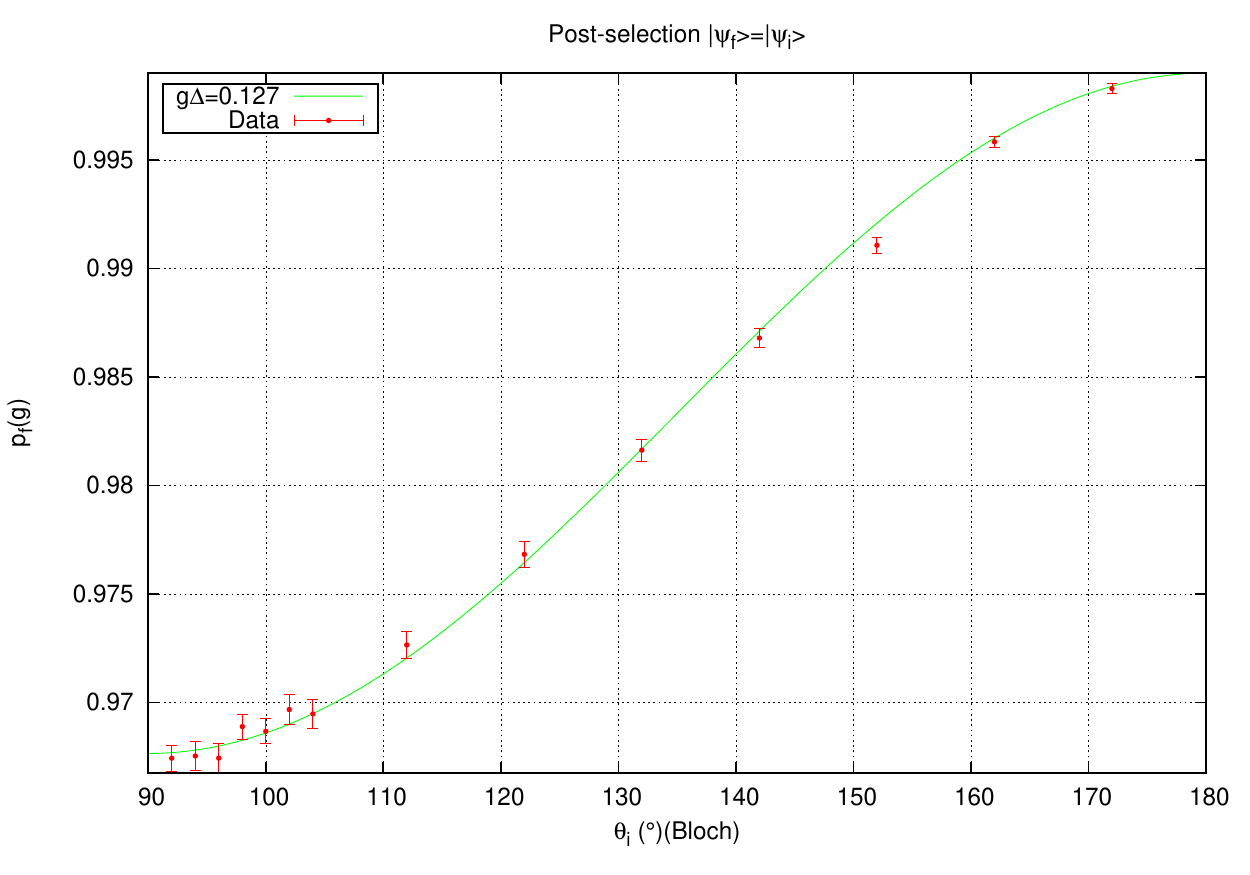}
\end{tabular}
\caption{Measured post-selection probability as a function of the initial polarization states. The curve joining the experimental data (dots) is a fit of the parameter $g\Delta$ for the given visibilities $\nu_0$ and $\nu_{\pi/2}$ measured experimentally before the interaction is applied. Due to the decoherence processes in the interferometer, the highest achievable information is at most about $50\%$ of the quantum information $F_Q$, in the region closer to the equator in the Bloch sphere ($\theta_i=\pi/2$), where the meter provides no useful information.}
\label{fig:plot-pf}
\end{figure}

\begin{figure*}[h]
\begin{tabular}{cc}
\includegraphics[scale=0.7]{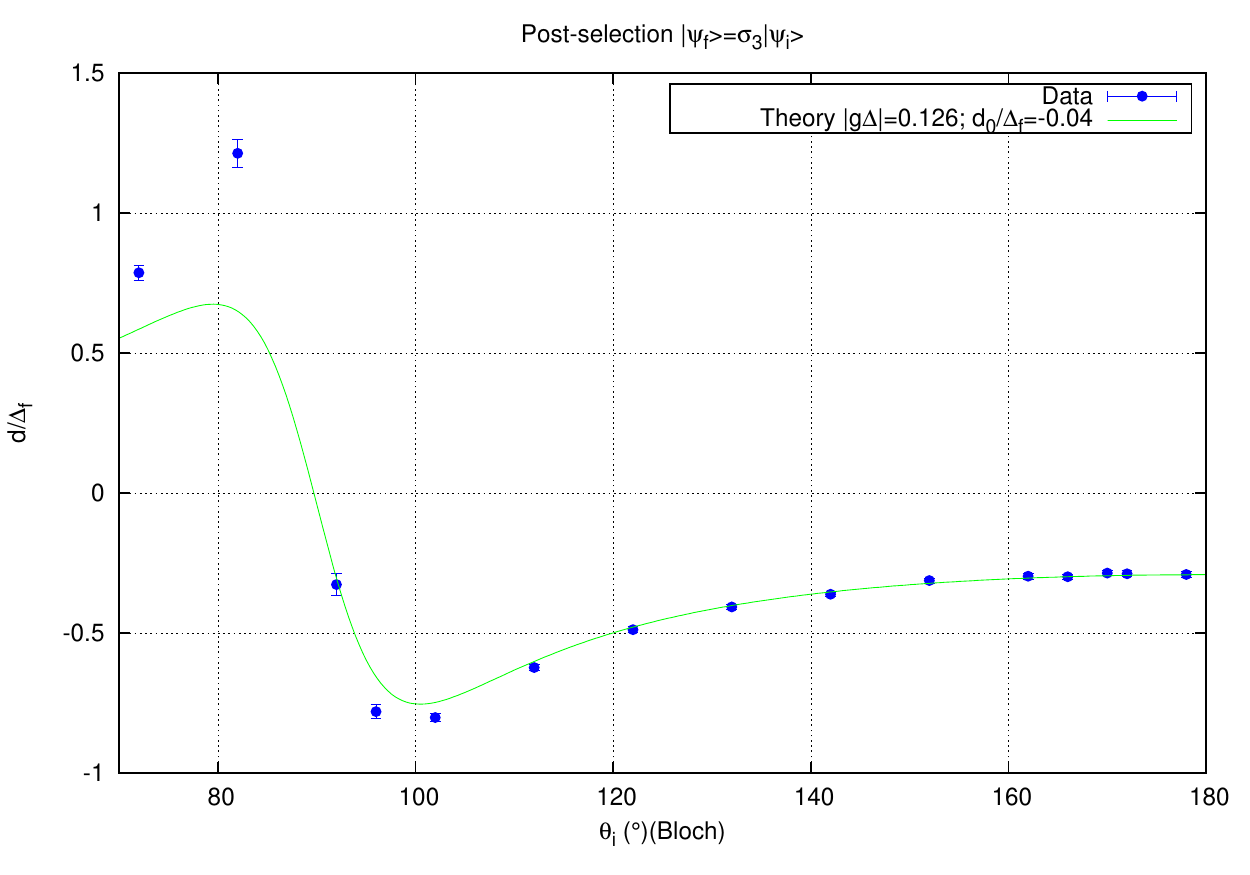}
& 
\includegraphics[scale=0.7]{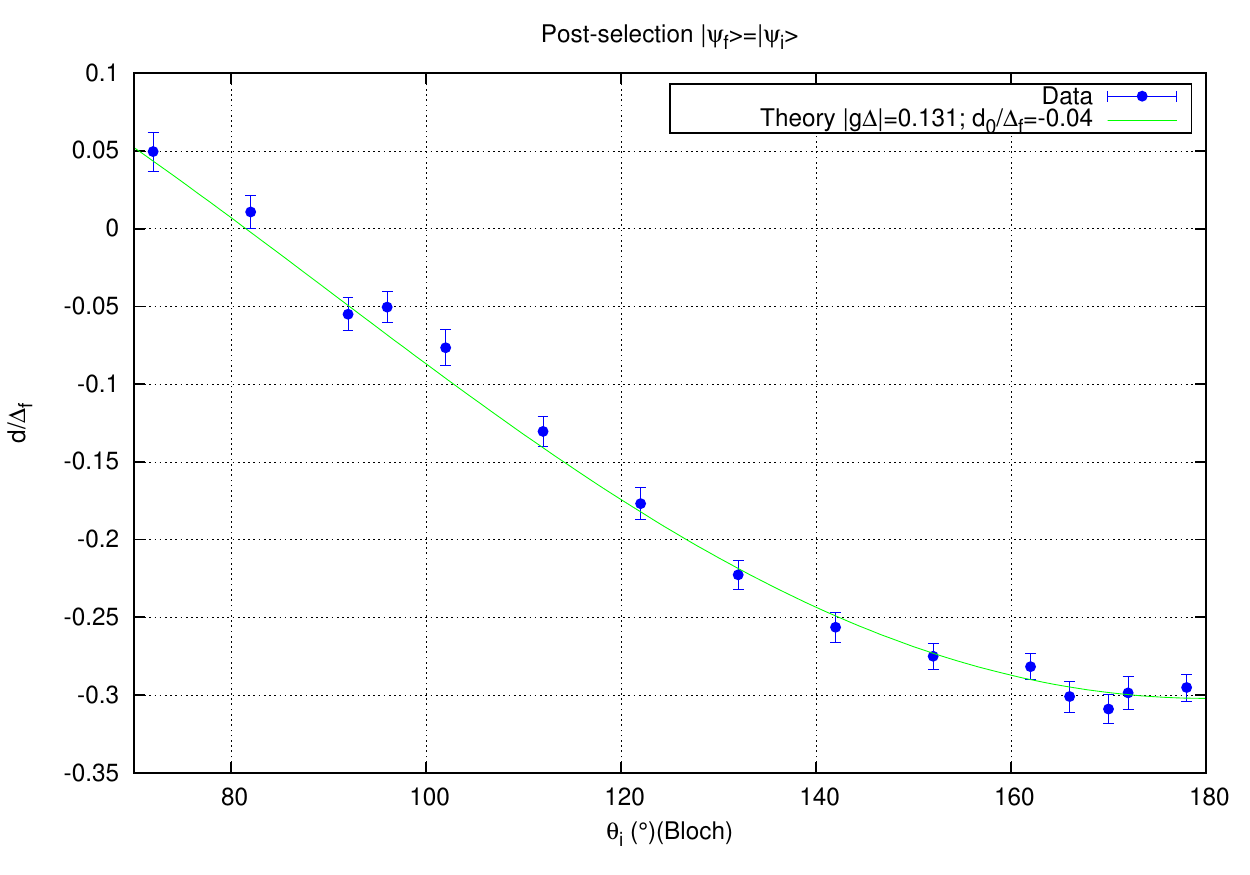}
\end{tabular}
\caption{Measured meter shift. Here we show the inferred shift of the gaussian wave-packet using the BBS technique for different initial polarization states in the Bloch sphere. It is worth noting that there is an amplification effect for the case $\ket{\psi_f}=\hat{\sigma_3}\ket{\psi_i}$, as seen from the graph. However, as the amplification gets stronger, the wavefunction gets distorted, and the BBS technique does not capture the correct shift (since it is meant to work for a gaussian wavepacket), as one can see from the points closer to the peaks. There is no amplification in the case $\ket{\psi_f}=\ket{\psi_i}$.}
\label{fig:desl-meter}
\end{figure*}

We now describe how the meter measurements are performed. Since the displacement is measured through the imbalance between the two halves in the transverse plane, one has to calibrate the detector before the interferometer is misaligned ($g=0$) to set the reference point. This is realized by matching the counts in the two halves of the detector (within statistical fluctuations of the photocounts) when the pre- and post-selected states are $\ket{+}$ (for which $\theta_i=\pi/2$), once this post-selection scheme is expected to have a null displacement according to Eq.\eqref{deslocamento-k-DD} for any value of the coupling $g$. After displacing the mirror M5, we are able to measure the new values of the intensities $N_L$ and $N_R$ by sequentially inserting and removing the BBS at the same position for the each post-selected state. However, the calibration was realized with a micrometer (10 micron precision, mounted on the BBS), which did not have the desired precision. We then added a constant to the theoretical displacement by simply replacing $d\mapsto d+d_0$, to account for any experimental error in the reference point, that best describe the data set. It is expected that this constant $d_0$ should be very small compared to the beam size at the focus, which were confirmed by our data (see Figures \ref{fig:plot-pf} and \ref{fig:desl-meter}).

\subsection{Experimental results}

Here we show the experimental results obtained for the post-selection probability (Eq.\eqref{pf-DD-0}) and the beam displacement (Eq.\eqref{deslocamento-k-DD}).

\newpage

\bibliographystyle{apsrev} %Style of Bibliography: unsrt / plain / apalike / amsalpha / ...
%\bibliography{references2}

%\clearpage
%
%\appendix
%\onecolumngrid
%
%\begin{center}
%{\Large Supplemental Material}
%\end{center}

\end{document}